\documentclass[12]{article}
\usepackage{amsmath}
\usepackage{color}

\definecolor{red}{rgb}{1,0,0}
\definecolor{blue}{rgb}{0,0,1}
\definecolor{green}{rgb}{0,1,0}
\definecolor{black}{rgb}{0,0,0}
\definecolor{yellow}{rgb}{1,1,0}
\definecolor{mdwblue}{rgb}{0.2,0.2,0.6}
\definecolor{gray}{rgb}{0.7,0.7,0.7}
\definecolor{darkgreen}{rgb}{0.2,0.7,0.2}
\begin{document}
\title{Towards a fully consistent relativistic quantum mechanics and a change of perspective on quantum gravity.} \author{Johan Noldus\footnote{Johan.Noldus@gmail.com }} \maketitle
\begin{abstract}
This paper can be seen as an exercise in how to adapt quantum
mechanics from a strict relativistic perspective while being
respectful and critical towards the experimental achievements of the
contemporary theory.  The result is a fully observer independent
relativistic quantum mechanics for $N$ particle systems without
tachyonic solutions.  A remaining worry for the moment is Bell's
theorem.
\end{abstract}
\section{Introduction}
As is often stated, quantum mechanics and general relativity are the
great accomplishments of the previous century.  However, their
unification is particularly troublesome because of the contradictory
axioms they are constructed from.  General relativity is a theory of
space time, reality, objectivity, locality of interaction \ldots
while quantum theory is about space plus time, predicting outcomes
of experiments, subjectivity, the instantaneous character of as well
measurement and interaction.  There has been an immense effort
during the last forty years in trying to force general relativity
into quantum theory with very little success.  In this paper, we
boldly address the opposite task and reexamine quantum mechanics in
a Bohmian spirit from a strict, die hard relativistic point of view
and see how far it brings us.  The first obstacle we meet is the
multi particle Schroedinger equation and I shall discuss objections
against it in section (2) where also new multi particle equations
are derived which admit straightforward observer independent
relativistic generalizations. The key property is that the single
$N$ particle wave which lives on configuration space cross time is
replaced by $N$ single particle waves which live on \emph{space
time}.  Of course our new equations have the correct classical
limit. Moreover, interactions between the different waves are
mediated by classical gauge fields such as the electromagnetic four
vector which implies that there are no instantaneous processes
taking place which jeopardize causality.  The physical picture we
end up with is one of wave packages moving around in space time
containing one particle each. The wave packages do not spread faster
than with the speed of light and interact through fields which
satisfy hyperbolic equations of motion.  This weltanschaung is the
complete opposite of the one employed in quantum field theory which
deals with wave functions on infinite dimensional configuration
space. In my opinion, a covariant set of field equations which unite
gravitation which quantum mechanics and matter is more in our reach
now and I shall present some ideas of how to proceed in that
direction in the epilogue.
\\*
\\* In section $3$, I shall present an elegant solution to the de Broglie mass problem which was
already mentioned in the literature \cite{Shojai}.  Furthermore, in
section (3.2) I discuss the same objections against the $N$ particle
Schroedinger equation from a relativistic point of view and present
the appropriate relativistic equations.  The conclusion from the
latter exercise is that it is possible to slightly modify quantum
mechanics into an objective, realistic and causal theory.  However,
as good relativists, we obviously reject spooky action at a distance
and therefore subjective entangled states.  The price to pay is the
rejection of quantum mechanics and some of the
\emph{interpretations} of the measurement results which contain some
trivial hidden assumptions which might not be as innocent as they
seem.  Also, it might be that the assumptions behind some of the
Bell inequalities were too naive and this is discussed a bit more at
the end of section (3.2). As a matter of notational convenience,
Greek letters such as $\alpha, \beta$ are space indices while $\mu$
and $\nu$ denote space time indices.  The signature of the metric is
taken to be $+ - - -$ and $D$ is the dimension of space.
\section{Non relativistic Bohm - de Broglie quantum mechanics}
In this section, I present the main concepts of Bohmian QM and shall
not refrain from going into a rather detailed interpretation.  The
latter which, in my opinion, was the cause of all confusion.  Let us
start with the one particle case in $R^D$, the Schroedinger equation
reads:
\begin{equation} \label{1}
 i \hbar \partial_{t} \Psi (x^{\alpha},t) = - \frac{\hbar^{2}}{2m} \Delta
\Psi (x^{\alpha},t) + V(x^{\beta},t) \Psi
(x^{\alpha},t).\end{equation} Going over to polar coordinates $\Psi
(x^{\alpha},t) = R(x^{\alpha},t) \exp{(iS(x^{\beta})/\hbar)}$,
(\ref{1}) becomes
\begin{eqnarray} \label{2}
\partial_{t} S + \frac{1}{2m} \left| \vec{\nabla} S
\right|^{2} + V - \frac{\hbar^{2}}{2m} \frac{\Delta R}{R} & = & 0 \\
\partial_{t} R^{2} + \frac{1}{m} \vec{\nabla} . \left(R^{2} \vec{\nabla}S
\right) & = & 0
\end{eqnarray}
The first equation is of the Hamilton Jacobi type with a dynamical,
nonlocal term $Q_{cl} = - \frac{\hbar^{2}}{2m} \frac{\Delta R}{R}$,
the so called quantum potential.  The second equation is a classical
continuity equation for the probability density $R^2$ provided the
fluid lines satisfy the differential equation:
\begin{equation} \label{4} \dot{x}^{\alpha} = \frac{1}{m} \partial^{\alpha} S. \end{equation} Now, it
is easy to check that \begin{equation} \label{5} m \ddot{x}^{\alpha}
= -
\partial^{\alpha} (V + Q) \end{equation} which is nothing but Newton's equation. \\* \\*  Hence, in Bohmian QM one extends reality by saying that the particle
\emph{has} a well defined position and momentum at each moment in
time; the probability density $R^2$ being nothing but a nonlocal
guidance mechanism for the particle.  This interpretation is
somewhat unsatisfying because of the role played by $Q_{cl}$ and we
shall improve upon this in section (3.1) where the latter shall be
given a space time geometrical meaning.  We continue this section by
elucidating the geometrical meaning of the argument field
$S(x^{\alpha},t)$, i.e. a small introduction into the geometry of
Hamilton Jacobi theory is given.
\\* \\* Let $\mathcal{L}(x^{\alpha},v^{\beta},t)$ be a time
dependent Lagrangian with corresponding functional
$$ I(\gamma,t) = \int_{t_{0}}^{t} \mathcal{L}(\gamma^{\alpha}(s), \dot{\gamma}^{\beta}(s),s) ds $$
on the space of curves $\gamma :[t_0 , t_1  ] \rightarrow R^{D} $
and $t_0 \leq t \leq t_1$.  The function $S(x^{\alpha},t)$
determines hypersurfaces in $R^{D+1}$ and we can ask how the
parameter $S(x^{\alpha},t) = \sigma $ changes with $dI(\gamma,t) =
\mathcal{L}(\gamma^{\alpha}(t), \dot{\gamma}^{\beta}(t),t) dt$.
Assuming that an initial surface $S(x^{\alpha},t) = \sigma_{0}$ has
been chosen so that $\frac{dt}{d\sigma}|_{\sigma_0} \neq 0$, we can
locally choose a sigma orientation such that
$$\frac{d\sigma}{dt} = \partial_{t} S(x^{\alpha},t) +
\partial_{\beta} S(x^{\alpha},t) \dot{x}^{\beta}
> 0.$$  Now, given $d\sigma$ fixed, the curve $(\gamma^{\alpha}(t),t)$ in $R^{D+1}$ shifts by
$(\frac{d\gamma^{\alpha}(t)}{dt}\frac{dt}{d\sigma},\frac{dt}{d\sigma})d\sigma$.
Therefore $\frac{d\gamma^{\alpha}(t)}{dt}$ is invariant under
reparametrisations of $\sigma$ and we can meaningfully ask in which
directions $\frac{dI(\gamma^{\alpha}(t),t)}{d\sigma}$ is extremal,
that is:
$$\partial_{\dot{x}^{\alpha}} \left( \frac{dI(\gamma^{\alpha}(t),t)}{d\sigma} \right) = 0.$$
An elementary calculation reveals that \begin{equation} \label{3}
\partial_{x^{\alpha}} S(\gamma^{\beta}(t),t) =
\frac{\Delta}{\mathcal{L}(\gamma^{\beta}(t),\dot{\gamma}^{\beta}(t),t)}
p_{\alpha} \end{equation} where $\Delta =
\frac{d\sigma}{dt}(\gamma^{\beta}(t),\dot{\gamma}^{\beta}(t),t)$,
which is homogeneous in first degree in the velocities, and
$p_{\alpha} = \frac{\partial \mathcal{L}}{\partial
\dot{x}^{\alpha}}$.  Now, it is an elementary exercise to show that
$S(x^{\alpha},t)$ satisfies the Hamilton Jacobi equation if and only
if $\Delta(\gamma^{\beta}(t),\dot{\gamma}^{\beta}(t),t) =
\mathcal{L}(\gamma^{\beta}(t),\dot{\gamma}^{\beta}(t),t)$ for any
member $\gamma$ of the congruence determined by equation (\ref{3}).
The latter implies that any action integral of any member of the
congruence between $S(x^{\alpha},t) = \sigma_i$ equals $\sigma_2 -
\sigma_1$.  The latter is inviting one to call the surfaces
$S(x^{\alpha},t)$ satisfying the Hamilton Jacobi equation to be
\emph{geodetically} equidistant.  Moreover, it is fairly elementary
to show that $S$ satisfies the Hamilton Jacobi equation if and only
if any member of its congruence satisfies the Euler Lagrange
equations.  The reader can find a more extensive treatment in
\cite{Butterfield}
\\*
\\* Historically, Schroedinger knew that the Hamilton Jacobi
function determines a family of surfaces which are geodetically
equidistant and crossed orthogonally by the specified congruence of
solutions to Hamilton's equations \cite{de Broglie}. Therefore, it
appeared natural to him that these surfaces correspond to the phase
factor of a \emph{monochromatic} wave which consequently has to
satisfy the Schroedinger equation; $\hbar$ being needed for
dimensional reasons only.  In those days, one was aware of the
wave-particle aspect of any \emph{substance} and since monochromatic
waves would not be physically adequate (substance would interfere
periodically at arbitrary large spatial separations) it deemed
necessary to include an amplitude.  Scaling invariance of the
Schroedinger equation with respect to this amplitude hinders it from
having a \emph{physical} meaning and consequently a probability
interpretation becomes natural.  This is all very well for
\emph{one} particle: one started from a natural idea inspired by
classical mechanics, saw that it needed extension in order to
correspond to localized substance in reality and then became
immersed in the casino game.  Moreover, at this stage there is no
real conflict with Einstein's principles of locality and causality
since it \emph{is} possible to give a geometrical meaning to the one
particle wave function assuming a Bohmian point of view.  This can
be realized by taking a Weyl geometrical approach as followed by
Santamato, leaving the relativistic Klein Gordon equations intact
\cite{Santamoto1} \cite{Santamato2}. He writes down an action
principle for a relativistic particle and a connection in an
external electromagnetic field and background space time.  It turns
out that the connection $\Gamma$ is semi-metrical and that the gauge
freedom in $\Gamma$ (which determines the curvature $R$ of the
connection in a flat background) is dynamically determined such that
$R$ equals the quantum potential.  Another approach consists in
asserting that quantum mechanics \emph{really} changes the conformal
factor of the space time metric which results in a modification of
the continuity equation for the probability density in $D>1$.  \\*
\\*  In my personal opinion, quantum mechanics went in the wrong direction when Schroedinger wrote down his equations for a system of $N$
particles.  In doing so, he started from the Hamilton Jacobi
equations for the \emph{composite} system of particles, interacting
\emph{instantaneously} through global forces.  Thereby, he
completely ignored the lesson Einstein presented in his theory of
general relativity that interaction occurs causally and moreover is
mediated by fields satisfying hyperbolic partial differential
equations, a comment which was also made by Louis de Broglie
\cite{de Broglie} (p. 140). Another valuable lesson of that theory
was that a conserved \emph{energy} has only a global meaning in a
static, asymptotically flat space time and therefore is an highly
inappropriate concept to start from. Inevitably, this brought
Schroedinger to the conclusion that the wave function has to live in
\emph{configuration} space cross some time, instead of in real
physical space time, making it devoid of any realistic
interpretation\footnote{Bohm, quite incomprehensively, does give a
realistic meaning to this wave!}.  However, what is not well
stressed at all in the standard literature, is that the
\emph{probabilistic} interpretation of the multi particle wave
function is quite unsatisfying too.  If I were a real Newtonian, I
would demand that the fact that the probability of having some
particle at time $t$ present is one, is \emph{independent} of the
consideration of other particles.  This is in conflict with the
contemporary dogma which only allows a particle to exist in space
\emph{given} that all other particles do.  Therefore, it seems to me
that each particle should have its own wave function with the
corresponding continuity equation.  In the remaining part of this
paper, I will gradually restore a \emph{realistic} interpretation of
the $N$ particle quantum mechanical system and, as highlighted in
section (3.2), such enterprise is intimately connected to the
principle of relativistic invariance.  The first step consists in
recognizing that indeed each particle has its own wave function
living in space time and that the latter are interacting causally
through gauge fields.  Next, a natural space time geometrical
meaning is given to the quantal guidance mechanism and this is
discussed in more detail in section (3.1).  Therefore, the only
probabilistic remnant in the theory concerns the unknown initial
positions of the particles.  The measurement process is, in this
context, a multi particle (scattering) problem and deeper insight
into this subject will require a lot of future computational work.
\\*
\\*  From now on, the reader should refrain from any Newtonian thought and reason as a proper relativist
(or at least switch off his/her Galilean preconceptions).  As a
warmup, consider a system of two particles whose wave functions
$\Psi_i(x^{\alpha},t_0)$ have disjoint support and spreads $\Delta
E_i$, $\Delta x_{i}^{\alpha}$ of the order $\sqrt{\hbar}$ around
center values $\tilde{E}_i$ and $\tilde{x}_{i}^{\alpha}$
respectively.  Although the two particles might interact after we
have prepared the system, it is very reasonable to think that they
are independent initially\footnote{Although, ardent quantum
physicists of the environmental decoherence clan might question
this.  However, I have never seen any detailed \emph{microscopic}
model for long range entanglement between the subsystem under study
and the environment. One typically proposes, however, the most
convenient possible effective interaction Hamiltonian.  Moreover,
this attitude is highly counterintuitive since one would expect the
gun of particle $2$ to shelter it from any long range influence from
particle $1$ as well as to destroy any preexisting correlations.}.
Therefore, we \emph{can} speak of the separate energies of the
particles (and they could be determined after some time by letting
the particles hit a screen and by measuring the impact factors)
which logically implies the need for \emph{two} Galilean time
parameters.  One can formalize this intuition by studying the
classical problem of $N$ particles with charge $e_j$ and mass $m_j$
interacting through the Coulomb force. In this case, the Hamiltonian
is:
$$\mathcal{H}(x^{\alpha_1}_1, \ldots, x^{\alpha_N}_N, p_{\alpha_1}^1, \ldots , p_{\alpha_N}^N) = \sum_{i=1}^{N} \frac{p_i^2}{2 m_i} + \sum_{i < j}
\frac{\alpha e_ie_j}{r_{ij}} $$ with $\alpha$ a positive coupling
constant. However, this theory neglects the facts that interactions
between the different particles occur with finite speed and that the
couplings between the particles and the fields generated by the
others are local as well.  Taking into account the latter
considerations results in the following correct theory:
\begin{eqnarray*}
\mathcal{H}_{i}(x^{\alpha}_i,p_{\alpha}^i,t) & = & \frac{ | p^{i} -
\frac{e_i}{c} \sum_{j} A_{j}(x^{\alpha}_i,t)|^2}{2 m_i} + e_i
\sum_{j} \phi_j(x^{\alpha}_i,t) \\
\partial_{\mu} F^{i \mu \nu}(x^{\alpha},t) & = & 4 \pi c^{-1} J_{i}^{\nu}(x^{\alpha},t) \\
\partial_{\mu} J_{i}^{\mu}(x^{\alpha},t) & = & 0
\end{eqnarray*}
where the $\phi_j$ and the $A_{j}^{\alpha}$ are the field and vector
potentials respectively, $A^{j}_{\mu} = \left( \phi_j , A_{j}
\right)$ is the $j$'th gauge four vector and $F^{j}_{\mu \nu} =
\partial_{j [ \mu} A^{j}_{\nu ]}$ is the $j$'th field strength.
The $N$ Hamilton functions enable one to write down the orbits of
the particles as explicit functions of time and the electromagnetic
fields of the other particles, given the initial positions and
momenta, i.e. $x^{\alpha}_i(t) \equiv x_{i \, \{A_{j}^{\mu}
\}}^{\alpha}(t)$.  However, strictly speaking, the four currents are
$$\left( J_{i}^{\mu}(x^{\alpha},t) \right) = e_i \left(c ,
\frac{p^{i}(t) - \frac{e_i}{c} \sum_{j}
A_{j}(x^{\alpha}_i(t),t)}{m_i} \right) \delta^{(3)}\left( x^{\alpha}
- x^{\alpha}_i(t) \right) $$ and these obviously satisfy the
conservation laws in the distributional sense on shell (that is when
the Hamilton equations are satisfied).  Now, it becomes very clear
why the j'th particle determines its own wave function satisfying a
one particle Schroedinger equation with a time dependent
Hamiltonian. Therefore, the new $N$ particle quantum mechanical
equations read:
\begin{eqnarray*}
\partial_{t_j} S_j(x_{j}^{\alpha},t_j) + e_j \sum_{i} \phi_i(x_{j}^{\alpha},t_j)  - \frac{\hbar^2}{2 m_j} \frac{\Delta_j
R_j(x_{j}^{\alpha},t_j)}{R_j(x_{j}^{\alpha},t_j)} + & & \\
\frac{\left| \vec{\nabla}_j S_j(x_{j}^{\alpha},t_j) - \frac{e_j}{c}
\sum_{i} A_{i}(x_{j}^{\alpha},t_j) \right|^2}{2 m_j}
 & = & 0 \\
\partial_{t_j} R_{j}^2(x_{j}^{\alpha},t_j) + \frac{\vec{\nabla}_j .
\left( R_{j}^2(x_{j}^{\alpha},t_j) \left( \vec{\nabla}_j
S_j(x_{j}^{\alpha},t_j) - \frac{e_j}{c}  \sum_{i}
A_{i}(x_{j}^{\alpha},t_j) \right) \right)}{m_j} & = & 0  \\
\dot{x}_{j}^{\alpha} - \frac{\vec{\nabla}_j S_j(x_{j}^{\alpha},t_j)
- \frac{e_j}{c}  \sum_{i} A_{i}(x_{j}^{\alpha},t_j)}{m_j} & = & 0
\end{eqnarray*}
We still need $N$ Maxwell's equations for the electromagnetic
fields, that is we need to identify the correct electric currents.
Obviously, the probability currents determine the charge currents
with charge densities $\rho_j = e_j R^{2}_{j}$ and spatial currents
$\boldmath{\vec{j}}_{j} = e_j R_{j}^{2} \frac{\vec{\nabla}_j
S_j(x_{j}^{\alpha},t_j) - \frac{e_j}{c} \sum_{i}
A_{i}(x_{j}^{\alpha},t_j)}{m_j}$.  The latter is equivalent to
stating that the \emph{possibility} of the particle being at some
place influences the dynamics of the other particles and the quantum
mechanical continuity equations imply that $\partial_{\mu}
J_{j}^{\mu} = 0$.  Therefore, the remaining equations are (in cgs
units):
$$ \partial_{\mu} F^{j \mu \nu} = 4 \pi c^{-1} J_{j}^{\nu}. $$
 The above system obviously produces good Newtonian laws with the
correct classical limits.  Restricting to the two particle case, it
becomes possible to ask questions like: given the fact that particle
one passed at time $t_1$ through the region $\mathcal{O}_1$ of
space, what is the probability that particle two shall pass through
$\mathcal{O}_2$ at a later time $t_2
> t_1$?  Or, assuming that particle one shall pass through
$\mathcal{O}_1$ in five seconds from now, what is the probability
that the second particle goes through $\mathcal{O}_2$ now?
Obviously, performing an experiment to check the above assertions
will change the particles momentum and therefore the outcome of the
result, but that does not invalidate the possibility for addressing
such questions.  So in this interpretation, God does still play dice
but in real physical space \emph{time} and not in configuration
space. \subsection{Comments} Verschelde kindly pointed out to me
that the formulae I end up with coincide - in the case of
\emph{vanishing} vector potentials - with those of Hartree theory,
the latter being known not to give correct results for the Helium
atom\footnote{In case the vector potentials do not vanish, one
typically resorts to an operator formalism for the latter which
coincides with my ansatz in the first order when doing perturbation
theory around the stationary solution.}. Does this mean that (a) my
reasoning is abundant and (b) one of my assumptions is incorrect?
The answer to both arguments is clearly no for me, since (a) all
derivations of Hartree assume the multi particle Schroedinger
equation to be correct and set up a global variational principle to
\emph{approximate} its solution by a product state while I simply
get my formulas in a much more elegant way by \emph{rejecting} the N
particle equation.  Concerning (b): let me note that spin is not
included yet in my theory, something which is likely to be done by
making use of the ideas behind Einstein Cartan theory instead of
spinors. One reason for this is that anti-symmetrization is not
going to lower the energy for the Helium atom since the lowering
terms in Hatree Fock only involve wave functions of equal spin.
Moreover, anti-symmetrization would not be allowed in my theory a
priori since we have previously rejected the notion of \emph{one}
wave function for $N$ particles. Further evidence against the
Pauli-Dirac description of spin is provided by the fact that the
\emph{spin statistics} theorem is essentially a \emph{kinematical}
result while one would intuitively expect it to be of a
\emph{dynamical} nature \cite{Sorkin}. Although the topological
observations involved concerning the swap operation on the space of
$\epsilon$-localized basis states - which is mathematically a
principal $U(1)$\footnote{When we ignore internal degrees of freedom
or demand that the latter trivially factorize. This so called
absolute recognizability constraint is to say the least very
questionable and not satisfied in most practical situations.} fiber
bundle over (non-relativistic) configuration space modulo
permutation operations corresponding to identical particles - are
cute, they are themselves questionable and have no real physical
underpinning\footnote{The treatment of the $U(1)$ factor is entirely
similar to the mathematics behind the Bohm Aharonov effect. In the
latter however, the topological defect is generated by the fact that
the vector potential - which dictates the coupling between the
particle's momentum and the electromagnetic field - cannot be
continuously defined on a simply connected domain in space.  As
such, this effect has a \emph{dynamical} reason while the spin
statistics theorem is entirely kinematical.}.  Therefore, we need to
find a spin dynamics which results in a repulsion of equally
spinning\footnote{Perhaps there should be attraction on very tiny
scales.} and attraction of opposite spinning particles. The energy
lowering effect in Helium could therefore be accomplished by a
specific anti-symmetric tensor field (a \emph{spin} field if you
want) which obeys the above dynamical rules and this brings us into
the realms of Einstein Cartan theory. Moreover, gravitation is
expected to indirectly play - say by gravitationally induced
electromagnetism - a significant role too.
\\* \\* So, the above result has to be interpreted as follows:
\emph{given} flat space time physics, taken into account that a
\emph{proper} form of spin has not been included yet and assuming
that the description of a particle by a mathematical point is fairly
accurate in these circumstances\footnote{Something which I strongly
disagree with.}, \emph{then} it follows that the above theory is the
preferred candidate from the relativistic point of view; the latter
which is derived from first principles in section (3.2). This, by
itself, is a worthwhile result in my opinion since it clearly
indicates the direction in which a (deterministic?) reformulation of
quantum mechanics could be found.
\section{Relativistic Bohm - de Broglie quantum mechanics for a zero spin particle revisited.}
In this section, we derive a proper nonlinear extension of the
classical Klein Gordon equation which has no tachyonic solutions for
one particle as well for multiple particles. \subsection{The single
particle} Polar decomposition of the Klein Gordon field
$\Psi(x^{\mu}) = R(x^{\mu}) \exp{\left(i S(x^{\mu})/\hbar \right)}$
gives the following pair of equations:
\begin{eqnarray}
\partial_{\mu} \left(R^2 \partial^{\mu}S \right) & = & 0 \\
\partial_{\mu}S \partial^{\mu}S = m^2 c^2 \left( 1 + Q_{rel} \right) & \equiv & \mathcal{M}^2 c^2
\end{eqnarray}
where $Q_{rel} = \frac{\hbar^2}{m^2 c^2} \frac{\partial_{\mu}
\partial^{\mu} R}{R}$ is the dimensionless relativistic quantum potential
and $ \partial_{\mu}
\partial^{\mu} = \frac{1}{c^2}
\partial^{2}_{t} - \Delta$ is the d'Alembertian.  Now, there exist
solutions of the KG equation for which $Q_{rel} < -1$ so that the
relativistic current defined by $$\frac{dx^{\mu}(\lambda)}{d\lambda}
= \frac{\partial^{\mu} S}{m}$$ gets spacelike as is illustrated in
the next example. \newtheorem{exie}{Example}
\begin{exie}
\end{exie}
For ease of notation, we shall work in the so called natural units
$\hbar = c = 1$ and examine the problem of one relativistic particle
moving in two dimensional Minkowski space time.  The family of
normalized wave packages we are interested in is given by:
$$\phi_{\epsilon}(x,t) = \frac{1}{\sqrt{\epsilon}\sqrt{1 + A^2}} \left( \int_{-\epsilon/2}^{\epsilon/2} \exp{(-i (\omega t - kx))} dk + A \int_{s - \epsilon/2}^{s + \epsilon/2}
\exp{(-i(\omega t - kx))} dk \right) $$ where $\omega^2 - k^2 = m^2$
and $A,s$ are real nonzero constants.  For any $L$ we can take
$\epsilon$ sufficiently small such that for all $(x,t) \in [-L,L]^2$
we have:
$$ \phi_{\epsilon}(x,t) \sim \frac{\sqrt{\epsilon}}{ \sqrt{1 + A^2}} \left( \exp{(-i m t)}  + A \exp{(-i(\omega t - sx))}
\right) $$ To calculate the particle's paths, note that with some
slight abuse of notation $S = -\frac{i}{2} \ln{\left(
\frac{\phi}{\overline{\phi}} \right)}$.  Then, $E = - \partial_{t}S
= \frac{\epsilon}{(1+A^2)R^2} \left( m + A^2 \omega + A(m + \omega)
\cos((m-\omega)t - sx) \right)$ and the momentum $P =
\frac{\epsilon}{(1 + A^2)R^2} As \left( A + \cos((m - \omega)t - sx)
\right)$.  Now the ratio $\frac{P}{E}$ equals:
$$ \frac{P}{E} = \frac{As \left( A + \cos((m - \omega)t - sx)
\right)}{\left( m + A^2 \omega + A(m + \omega) \cos((m-\omega)t -
sx) \right)}. $$  For $\cos((m - \omega)t - sx) = -1$ the above
expression reduces to:
$$ \frac{P}{E} = \frac{As(A - 1)}{ m + A^2 \omega - A(m + \omega)}. $$
For $A = \frac{2}{\omega}$, one obtains $\frac{P}{E} =
\frac{2s}{\omega(2 - m)}$.  Obviously, for $0 < m < 2$ and $\omega >
\frac{2\sqrt{m}}{\sqrt{4+m}}$ we have $\frac{P}{E} > 1$ which is
what we wanted to show.  \\*  \\* Restoring dimensions in the above
example makes one aware that, since $A$ has to be dimensionless, the
factor $2$ must have dimension of $1/s$.  Hence one needs an
invariant length, that is the Planck length $\l_p$.  Recall that in
standard units, $\l_p \sim 1,6. 10^{-35}$, $\hbar \sim 6,6.
10^{-34}$ and $c \sim 3. 10^{8}$.  Putting $A = \frac{2 \alpha
c}{\l_p \omega}$, where $\alpha$ is a dimensionless numerical
constant, leads to
$$ \frac{Pc}{E} = \frac{2 c^2 \l_{p}^{-1} \alpha s \hbar }{\omega(2c \l_{p}^{-1} \hbar \alpha - m c^2
)} $$ and we may safely assume that $m c^2 < c \l_{p}^{-1} \hbar
\alpha$ or $0 < m < \frac{\l_{p}^{-1} \hbar \alpha}{c} \sim (1,4)
\alpha. 10^{-7}$ kilo.  In order for the above expression to exceed
one, it is necessary that:
$$ \omega > \sqrt{\frac{4 m c^4 \l_{p}^{-2} \alpha^2}{\left(4 c \l_{p}^{-1} \hbar \alpha  - mc^2\right)}} $$
For $\alpha = 10^{-22}$  and $m = 10^{-30}$ which is about the
electron mass, the above estimate reduces to $\omega > 10^{21}$
Hertz, which is in the gamma ray frequency spectrum with wave vector
$s$ larger than about $10^{13}$.  All these numbers fall well within
the reach of conventional space time physics.  \\*
\\*
This result is making a consistent Bohmian interpretation of the
relativistic KG equation impossible.  However, there is no good
reason why the KG equation should be the physically correct one,
i.e. it might very well be that the equation requires non linear
corrections. This is not strange at all, since it is well known that
the superposition principle in -say- an environmental decoherence
interpretation of QM is responsible for the measurement
problem\footnote{In this particular approach, one has the so called
pointer basis problem as well as the problem of Poincar\'e
recurrence times.}. The latter problem is solved in a Bohmian
interpretation and therefore one can take the point of view that a
correct relativistic equation should be Poincar\'e invariant, have a
probability interpretation (scaling invariance with respect to $R$)
and no acausal Bohmian trajectories.  The most obvious ansatz which
satisfies all these requirements (and does not affect the classical
limit) is given by $\mathcal{M}^2 = m^2 \exp{(Q_{rel})}$ instead of
$m^2(1 + Q_{rel})$. Actually, the latter formula is suggested by the
following series of arguments\footnote{I have no conclusive proof
here.}. Equation (8) can be rewritten as
$$ \frac{dx^{\mu}(\tau)}{d\tau} =
\frac{\partial^{\mu}S(x^{\nu}(\tau))}{\mathcal{M}} $$ where the
parameter $\tau$ yields the eigentime of the congruence.  Direct
calculation yields: \begin{equation} \label{6}
\mathcal{M}(x^{\gamma}(\tau)) \frac{d^{2}x^{\mu}(\tau)}{d\tau^{2}} =
\left( c^2 \eta^{\mu \nu} -
\frac{dx^{\mu}(\tau)}{d\tau}\frac{dx^{\nu}(\tau)}{d\tau} \right)
\partial_{\nu} \mathcal{M}(x^{\gamma}(\tau)). \end{equation}  Now, we study the
non-relativistic limit of the above equation.  That is: $\tau \equiv
t$, where $t$ is the time coordinate associated to a freely falling
congruence of observers, $(x^{\mu}(t)) = (ct, x^{\alpha}(t))$ with
$| \dot{x}^{\alpha} | \ll c$.  The geodesic equation (\ref{6}) is
decomposed as follows:
\begin{eqnarray*}
\mathcal{O}\left(\frac{\dot{x}^{\alpha}(t)}{c}\right)^2 & = & c
\dot{x}^{\alpha} \partial_{\alpha} \mathcal{M} \\
\mathcal{M} \ddot{x}^{\alpha} & = & c^2 \partial^{\alpha}
\mathcal{M} + \dot{x}^{\alpha} \dot{x}^{\beta} \partial_{\beta}
\mathcal{M} - \dot{x}^{\alpha} \partial_t \mathcal{M}
\end{eqnarray*}
The second equation has only a good non relativistic limit provided
that $\partial_{\beta} \mathcal{M} \sim \frac{\mathcal{M}}{c^2}$
which implies that the first equation is identically satisfied.
Because of the former equation, it is reasonable to think that
$\partial_t \mathcal{M} \sim \frac{\mathcal{M}}{c^2}$ and therefore
$$ \mathcal{M} \ddot{x}^{\alpha} =  c^2 \partial^{\alpha}
\mathcal{M} .$$  Recalling the non-relativistic equation implies:
$$ m \ddot{x}^{\alpha} = - \partial_{\alpha}\left( m c^2 \ln \left( \frac{\mathcal{M}}{\mu} \right) \right) = - \partial_{\alpha} Q_{cl} $$
where $\mu$ is any mass scale.  This latter implies that
$\mathcal{M} = m \exp{\left( \frac{Q_{cl}}{m c^2}  \right)}$ which
suggests that the relativistic mass satisfies $\mathcal{M} = m
\exp{\left( \frac{Q}{2} \right)}$.  An important aspect is that
equation (\ref{6}) is the \emph{geodesic} equation associated to the
space time metric $g_{\mu \nu} = \frac{\mathcal{M}^2}{m^2} \eta_{\mu
\nu}$, a remark which seems to go back to de Broglie.  Since in the
Bohmian interpretation, a point particle is localized somewhere
within the wave package, it seems unnatural to attribute a mass
field $\mathcal{M}$ to it.  Moreover, the fact that equation (8) is
equivalent to
$$ g^{\mu \nu} \partial_{\mu} S \partial_{\nu} S = m^2 c^2 $$
strongly suggests that one \emph{really} should see the quantum
effects on the motion of the relativistic particle as being given by
the above conformal transformation of the space time metric.
Consequently, the continuity equation should be modified\footnote{A
simple calculation reveals that $\nabla_{\mu} ( R^2 \nabla^{\mu} S )
= \frac{m^2}{\mathcal{M}^2} \partial_{\mu} ( R^2
\partial^{\mu} S ) + \frac{m^2}{\mathcal{M}^3}(D-1) R^2
\partial_{\mu} \mathcal{M} \partial^{\mu} S$.  Therefore the continuity equation changes for $D>1$.} to
$$ \nabla_{\nu} ( R^2 \nabla^{\nu} S ) = 0$$ where
$\nabla$ is the metric connection associated to $g_{\mu \nu}$ and,
moreover $$\frac{dx^{\mu}(\tau)}{d\tau} = \frac{\nabla^{\mu} S}{m}
.$$  The latter considerations bring us to the anticipated
conclusion that one can attribute a natural space time geometrical
meaning to the non local quantal guidance mechanism which takes away
its mystery.  That is: the wave character of the particle deforms
space time geometry in such a way that the geodesic pattern of point
particles equals the extremely chaotic congruence of curves in
Bohm's theory.  The only remaining uncertainty being the initial
position of the particle.
\\*
\\* It is unfortunately not well known that the one particle Klein
Gordon theory as enunciated above has a consistent interpretation
\cite{Santamato2}. The theory is time reversal invariant which
allows for particles to travel the same space time curves but in
opposite time directions. It is the choice of time of the observer
which determines the notion of particle and antiparticle. Obviously,
this can give rise to negative probabilities but this is \emph{not}
a problem as explained in \cite{Santamato2}.
\subsection{Multiple particles}
In this section, I approach the objections in section (2) from a
relativistic point of view and show that the standard formalism is
incomplete. The same comment applies anyway to any approach aiming
to solve the Unitary/Reduction problem in quantum mechanics, without
addressing the issue of \emph{local} coordinate (\emph{active}
Lorentz) invariance of the quantum mechanical equations in a curved
(flat) background properly.  For simplicity, I present the equations
in the $D+1$ dimensional special relativistic case; the extension to
a curved background being straightforward.  In relativity,
configuration space is $R^{N(D+1)}$ (!) and it is therefore natural
to demand the wave function $\Psi$ to depend on the $N$ tuple of
\emph{space time} coordinates (unlike what is done in quantum field
theory where a split of space and time is made from the start), i.e.
$\Psi(x_{1}^{\mu_1}, \ldots , x_{N}^{\mu_ N})$. Then, the $N$
particle relativistic equations in Minkowski \emph{would} look as
follows:
\begin{eqnarray}
\sum_{j = 1}^{N} \frac{1}{m_j} \partial_{j \mu} S \partial_{j}^{\mu}
S & = & \sum_{j = 1}^{N} \frac{\hbar^2}{m_j} \frac{ \partial_{j \mu}
\partial_{j}^{\mu} R}{R} + \sum_{j = 1}^{N} m_j c^2 \\
\sum_{j = 1}^{N} \partial_{j \mu} \left(  \frac{R^2
\partial_{j}^{\mu} S}{m_j} \right) & = & 0
\end{eqnarray}
where $\partial_{j \mu}$ denotes the partial derivative to the
$\mu$'th coordinate of the $j$'th particle.  The latter equation is
again the usual continuity equation in configuration space.  Both
equations being second order hyperbolic partial differential
equations in $R$ and $S$.  In the above equations, each particle is
allowed to have its own independent Lorentz \emph{coordinate} system
and unless we fix a gauge by specifying the relative boost and
rotation factors between the different coordinate systems, the above
system leaves $2(N-1)$ variables undetermined\footnote{Even in case
$R_0, S_0$ and the first time derivatives would be equal to a
product, respectively a sum of one particle initial conditions
(expressed in the different local Lorentz coordinate systems), there
would not exist a unique solution.  In the specific case of a two
particle system, this is reflected in the existence of a free mass
parameter. Obviously, for entangled (in the standard meaning of the
word) initial conditions, the situation is far worse.}. The point is
that from a relativistic perspective, one should not know these
relative Lorentz transformations in order to find a unique solution.
This becomes very clear when one imagines the $N$ particles to be
initially spatially separated meaning that the initial $N$ particle
wave function is the product of $N$ one particle wave functions
whose supports are a distance $> \epsilon$ apart in the Riemannian
metric determined by the initial hypersurface.  In this setup, it is
completely irrelevant to even wonder about the relative Lorentz
transformations since in case interactions are present, they are
mediated by gauge fields and therefore the \emph{coupling} to other
particles is of a local nature.
 This argument becomes even more compelling in the context of true local coordinate transformations in a fixed
curved background space time where, in case transition functions
would exist, nothing is said about that part of the coordinate
patches which have no overlap anyway\footnote{Note that coordinate
transformations in a curved background have nothing to do with
diffeomorphism invariance as is often mistakingly quoted. The
correct behavior under coordinate transformations is a property of
tensors while diffeomorphism invariance is a dynamical property of
spacetime.}.
\\*  \\* In quantum field theory, a very special form of the above
discussed gauge fixing is accomplished by setting the relative boost
and rotation factors to zero, i.e. a globally consistent coordinate
system is chosen.  In order to avoid all possible confusion, let me
stress that this kind of gauge freedom should not be mixed up with
the so called local Lorentz invariance which lives on the
\emph{tangent bundle}; a symmetry the above equations obviously
posses also.  Moreover, the \emph{active} Lorentz invariance of QFT
(independence of the observer) is not the same as the above
coordinate symmetry, since the relative boost and rotation factors
remain null.  In the latter context, it is well known that the
necessity of a \emph{globally well defined} observer for a
\emph{fundamental} equation of physics to be written down becomes a
highly problematic matter, in particular in the standard
interpretation of quantum gravity where the status of the observer
him/herself gets jeopardized.  For all the above reasons, it seems
very natural and necessary to me that the laws of physics are
observer independent and that the well known problems that observers
bring along are indeed problems (of global geometrical or
topological nature) of the observers themselves and not of
fundamental physics.
\\*  \\*  Since standard quantum mechanics is shown to be incomplete, we need to
add $2(N-1)$ hyperbolic equations to the above system.  Taken into
account the results of section (2) and the discussion about the
positivity of $\mathcal{M}^2$ in section (3.1), the extension to the
relativistic case becomes obvious and the result reads:
\begin{eqnarray*} \left(
\partial_{j \mu} S_j(x^{\nu}_j) + \frac{e_j}{c} \sum_{i} A_{i \mu} (x^{\nu}_j) \right)
\left( \partial_{j}^{\mu} S_j(x^{\nu}_j) + \frac{e_j}{c} \sum_{i}
A_{i}^{ \mu} (x^{\nu}_j) \right) & = &  \mathcal{M}_{j}^{2}(x
_{j}^{\nu}) c^2  \\
\partial_{j \mu} \left( \frac{R_{j}^2(x^{\nu}_j)
\left( \partial_{j}^{\mu} S_j(x^{\nu}_j) + \frac{e_j}{c} \sum_{i} A_{i}^{ \mu} (x^{\nu}_j) \right) }{m_j} \right) & = & 0 \\
\partial_{j}^{\gamma} F^{j}_{\gamma \mu}(x^{\nu}_j) & = & 4 \pi c^{-1}
J^{j}_{\mu}(x^{\nu}_j) \\
\frac{dx_{j}^{\mu}(\tau_j)}{d\tau_j} - \frac{\left(
\partial_{j}^{ \mu} S_j(x^{\nu}_j) + \frac{e_j}{c} \sum_{i} A_{i}^{\mu} (x^{\nu}_j)
\right)}{\mathcal{M}_j} & = & 0
 \end{eqnarray*}  where $\mathcal{M}_{j}^{2}(x^{\nu}_j) = m_{j}^{2}
 \exp{ \left( \frac{\hbar^2}{m_j^2 c^2} \frac{\partial_{j \mu} \partial_{j}^{\mu}
 R_j (x^{\nu}_j)}{R_j (x^{\nu}_j)} \right) }$ and $$J^{j}_{\mu}(x^{\nu}_j)  =
 e_j R_{j}^2(x^{\nu}_j) \frac{\left( \partial_{j \mu} S_j(x^{\nu}_j) + \frac{e_j}{c} \sum_{i} A_{i \mu} (x^{\nu}_j)
\right)}{m_j}. $$ Again, one notices that the distributional charge
and mass densities are smeared out by the amplitude $R_j^2$ and
$\mathcal{M}_j$ respectively, which is different from the well known
result in quantum field theory where the dressed mass and charge are
still parameters and not fields.  A discussion about the
geometrization of the relativistic multi particle system will appear
elsewhere.  \\*
\\* In this section, we have shown that the principle of local
coordinate invariance forces one to extend standard quantum
mechanics.  In the latter extension, each particle has its own wave
package and interaction is mediated causally through
(electromagnetic) gauge fields. We used the modification of the
standard KG equation proposed in \cite{Shojai}, meaning that the
superposition principle is abandoned.  At this point, the reader may
wonder how it will be possible to describe entanglement between,
say, two spin $1/2$ particles as in the Einstein Podolsky Rosen
experiments since, in the standard description, this requires use of
states of the form $ \frac{1}{\sqrt{2}} \left( | \uparrow > |
\downarrow >  - | \downarrow > | \uparrow > \right)$.  Now one can
take two points of view: either the above state has no space time
meaning and belongs to a preordained Clifford bundle with the
canonically induced connection, or spin has a space time
interpretation.  Let me discuss the former point of view first: the
problem is that a change of spinor \emph{bundle} generically gives
rise to different physics which implies that the Dirac picture is a
subjective one, that is the left (right) up and down states do not
correspond to a physical reality of the left (right) particle, but
to the outcome of a so called spin (in reality a position)
measurement on it\footnote{This raises again the question about the
meaning of a spin observable whose existence is not, in any sense,
proven to be necessary. Likewise, one can question the need for a
momentum observable which is a (infinitely) delocalized concept and
therefore highly unphysical.}.  Therefore, from an objective point
of view, it is entirely meaningless to speak about a delocalized
state of total spin zero.  The same holds for the notion of angular
momentum since its very definition requires a global preferred
coordinate system, something which has generically no canonical
meaning.  Of course, one might dismiss these objections just on
practical grounds but I am afraid that such attitude is against any
serious attempt of unification and I shall grasp this opportunity to
take a strict relativistic point of view\footnote{In case the multi
particle Schroedinger equation were more criticized before, this
attitude would perhaps not be a rarity now.}. The only
relativistically sane explanation is that correlations in the
outcome of measurements are the consequence of interactions between
the wave packages (and perhaps both measurement apparatus) in their
common causal past (assuming that both particles have the same time
orientation!) and no singlet state should be needed to describe
that. Since we are aiming for an objective world view and rejected
measurement as it stands in the Copenhagen interpretation, the
standard explanation cannot be maintained here. Therefore, the
reason for the \emph{correct} amount of correlations in experiments
must be of an entirely different \emph{physical} nature. Hence, we
are forced to take the position that a particle simply \emph{has} a
definite spin vector which obeys a deterministic equation of motion
and that, in case of the EPR setup, it are merely the initial spin
vectors which have opposite directions\footnote{That is, a hidden
variable model is needed. Something which has always been maintained
by Einstein.}. As mentioned before, this cannot be realized within
the context of conventional Dirac theory and another set of
equations will be necessary. Let me also note that there is no
conflict with the so called Kochen Specker theorems: the fact that a
particle has classical properties does not imply that these can be
determined simultaneously through experiments, just as momentum and
position of a particle cannot be simultaneously measured in
classical physics\footnote{Even classically, a measurement of
position is going to influence the momentum of the particle.
Therefore, in my opinion, the Kochen Specker theorems deal with the
wrong question.}. However, I am contradicting some
\emph{interpretations} of experiments concerning Bell's theorem
\cite{Bell} (as did de Broglie and more recently 't
Hooft\cite{Hooft}) and there exists a vast literature about possible
loopholes in the relevant experiments which still allow for the
possibility of a local hidden variable theory. However, this is not
the place to enter into this subject in detail.  The reader will
most likely feel uncomfortable with my urge for reinstating
classical concepts, but let me point out that I merely came to these
conclusions by respecting well known relativistic principles.
Moreover, the possible states which can occur in nature are narrowed
down immensely due to the exclusion of delocalized multi particle
entangled states which I believe, is good news.
\\*
\\* The construction is this section is remarkably similar to the so
called \emph{u}-wave description in the double wave theory of de
Broglie \cite{de Broglie} as I found out later. He also recognized
some of the afore mentioned problems with the standard formalism but
did not want to give up the multi particle Schroedinger equation.
Instead, he forced himself to think of the $\Psi$ wave as an
entirely fictitious object, living on configuration space cross one
(!) time, which had only a probabilistic meaning.  I can only
imagine that the reason for his obstinate recognition of the
Schroedinger wave must have been rooted in a long tradition of
classical Newtonian mechanics\footnote{Not to speak about the
success of its application!  An argument which, even more today, is
supposed to kill all resistance. } which was only questioned by then
for about $10$ years and which is still today the prevalent view on
physical reality. This claim is supported by the fact that de
Broglie frequently speaks about three dimensional \emph{physical
space}, indicating the presumption of a globally preferred notion of
time.
\section{Conclusions} In this paper, I constructed a fully
consistent relativistic quantum theory of particles in which each
particle has its own wave function which lives in space time. It
remains to be seen how the theory could account for particle-anti
particle pair creation, this could in principle happen in case the
phase factors $S_j$ get singular.  At least, our theory has the
potentiality to make such process explicit or at least indicate in
which direction one should extend it.  This is in sharp contrast to
quantum field theory where this process happens instantaneously and
is put in by hand through the specific form of the Lagrangian
density.  Moreover, the theory is observer independent implying that
particles have an objective existence. Consequently: the distinction
particle - antiparticle and particle energy are just observer
dependent concepts which can be introduced later. Moreover, we have
\emph{deduced} by pretty airtight arguments that a (deterministic)
hidden variable model for spin is necessary, something Einstein has
maintained for most of his life.  The latter implies that somehow
the derivation of the Bell inequalities must be incomplete and I
believe that a careful reexamination is a necessary and extremely
worthwhile enterprise, taken into account the possible reward.  If
this turns out to be successful, then quantum mechanics is
compatible with the laws of general relativity and the objective of
a universal field theory suddenly seems a lot more realistic.  In
case any imaginable attempt to give a hidden variable explanation
for the experimental outcome would fail, then and only then would I
see myself obliged to resign from one of relativity's beautiful
principles.
\\*
\\*
Regarding this universal field theory, I see the program being
developed in two steps.  The first step consists in ignoring
particle spin for the moment, keeping the description of point
particles and writing down a geometrical action principle which
contains gravitation, electromagnetism and point matter.  I believe
we are prepared to take that step now and I have commented a few
times on how this could be done.  A second, and much more difficult
step, consists in probing the geometrical structure of an elementary
particle. I am entirely convinced of the idea that elementary
particles are just bound states of an interacting ensemble of a
fundamental substance of \emph{space time}\footnote{This in contrast
to string theory where elementary particles correspond to vibration
modes of a string which lives in a background space time.} and that
quantum numbers such as particle spin should be encoded into this
geometrical form. Therefore, the contemporary bundle descriptions of
external quantum numbers (how elegant they may be) are in my opinion
merely subjective pictures of a much deeper physical reality.
Insight into the latter subject would also throw more light upon the
\emph{how} of the particle scattering processes.
\\*
\\* Let me end by recalling the historical happening of $1949$,
where on the occasion of Einstein's $70$'th birthday quantum
physicists such as Born, Pauli and Heitler, members of the orthodox
school, shamelessly expressed in their articles their disappointment
at seeing Einstein persist in a negative attitude towards the purely
probabilistic approach to Wave Mechanics.  Perhaps people should be
more careful in making claims about issues which are not settled
yet, especially towards the great scientist who was the main actor
in both revolutions.
\section{Acknowledgements}
I would like to thank Lode Wylleman for carefully reading this
manuscript and for useful suggestions concerning its presentation.
Also, I am indebted to my old mentor Norbert Van den Bergh for
continuous encouragement and his insistence upon a locally causal
quantum mechanics.

\end{document}